\documentclass{article}
\date{}
\usepackage{graphicx}
\usepackage{amssymb}
\usepackage{amsmath}
\usepackage{amsthm}
\usepackage{amsfonts}
\usepackage{amssymb,amsbsy,amsmath,amsfonts,amssymb,amscd}
\def \Prob {{\bf P}}
\newcommand{\R}{\mathbb{R}}

\newcommand{\Z}{\mathbb{Z}}

\def \H{\mathbb{H}}
\def \N{\mathbb{N}}

\def \eps {{\varepsilon}}

\newtheorem {proposition}{Proposition}
\newtheorem {lemma}{Lemma}

\begin{document}

\title{
Conformal restriction,
highest-weight
representations and SLE}

\author{
Roland Friedrich  \footnote {Universit\'e Paris-Sud and IHES}
\ \
\and
Wendelin Werner
\footnote {Universit\'e Paris-Sud and Institut Universitaire de France}
}
\maketitle
%%%%%%%%%%%%%%%%%%%%%%%%%%%%%%%%%%%%%%%%%%%%%%%%%%%%%%%%%%%%
%%%  Abstract  %%%
%%%%%%%%%%%%%%%%%%
\begin{abstract}%
We show how to relate Schramm-Loewner Evolutions (SLE)
to highest-weight 
representations of infinite\--dimensional 
Lie algebras that are singular at level two, using the conformal restriction properties 
studied by Lawler, Schramm and Werner in \cite {LSWr}.
This confirms the prediction from conformal field theory that 
two-dimensional critical systems are related to  
degenerate representations.
\end{abstract}
%%%%%%%%%%%%%%%%%%%%%%%%%%%%%%%%%%%%%%%%%%%%%%%%%%%%%%%%%%%%
%%%  Main text (in English)  %%%
\section {Introduction}

The goal of this paper is to show how the Schramm-Loewner
evolutions (or Stochastic Loewner Evolutions, which is 
anyway abbreviated by SLE)
 can be used to interpret in a simple and
elementary way some of the starting points
of conformal field theory, stated by Belavin-Polyakov-Zamolodchikov
in their seminal paper \cite {BPZ}. In particular, we will see
how restriction properties studied in \cite {LSWr} can
be rephrased in terms of highest-weight representations of the
Lie algebra $\cal A$ of vector fields on the unit circle (and its
central extension, the Virasoro algebra).
The results in this paper were announced in the note \cite {FW}.

It is probably worthwhile to spend some lines
outlining our perception of the history of this subject
(see also the recent review paper by Cardy \cite {Ca3}):
It has been recognized by physicists some decades ago
 that two-dimensional systems from statistical physics near their
critical temperatures  have
some universal features. In particular, some
quantities (correlation length for instance) obey universal
power laws near the critical temperature,
and the value of the (critical) exponent in fact depends only on the
phenomenological features of the discrete system (for instance,
it is the same for the same model, taken on different lattices).
In order to identify the value of the exponents, two
techniques turned out to be very successful. The first one is
the ``Coulomb gas approach'' (see e.g. \cite {Ni} and the
references therein, as well as the reprinted papers in \cite
{ISZ}), which is based on explicit computations for some specific models.
The second one (see Polyakov \cite {Po},
Belavin-Polyakov-Zamolodchikov \cite {BPZ}, Cardy \cite {Ca1})
is conformal field theory.
Based on the analogy with some other problems, it is argued
in \cite {BPZ} that
two-dimensional critical systems are associated
to conformal fields. These fields should then satisfy certain
relations, such as the Ward identities, which then allow to
make a link with highest-weight representations of the Virasoro
algebra.
Then the critical exponents can be identified from the corresponding highest weights.

We now quote from \cite {Ka}: {\sl
``The remarkable link between the theory of highest-weight modules
over the Virasoro algebra and conformal field theory and statistical mechanics
was discovered by Belavin-Polyakov-Zamolodchikov \cite {BPZ1,BPZ}.
Conformal Field Theory has now become a huge field with ramifications
to other fields of mathematics and mathematical physics''.}
We refer
for instance to the introduction of
\cite {FbZ} and the compilation of papers in \cite {GO, ISZ}.
This approach has then been used to develop the related ``quantum gravity''
method (see e.g. \cite {Dqg}) and the references therein. 
 
It is worthwhile to stress some points:
The actual mathematical meaning, intuition or definition
of these fields (and their properties, such as the Ward identities)
in terms of the discrete two-dimensional models was to our knowledge never
clarified. Also, the notion of ``conformal invariance'' itself
for these systems remained rather obscure. In the case of
critical percolation, Aizenman \cite {Ai} formulated
clearly what it should mean, but for other famous models such as
self-avoiding walks, or Ising, the precise conjecture
 was never stated until recently.

In \cite {Ca2}, Cardy pointed out that in the case of critical percolation,
the arguments from \cite {BPZ,Ca1} could be used in order to
predict the exact formula for asymptotic crossing probabilities
of a topological rectangle by a percolation cluster.
This prediction was popularized in the mathematical community through the
review paper by Langlands-Pouliot-StAubin \cite {LPS}, that attracted
many mathematicians to this specific problem (including Stas Smirnov).
In that paper, the authors also explain how difficult it is
for mathematicians to understand Cardy's
arguments.

On a rigorous mathematical level, only limited progress towards the understanding
of 2D critical phenomena had been made before the
late 90's. In 1999, Oded Schramm \cite {S1}
defined a one-parameter family of random curves based
on Loewner's differential equation, SLE$_\kappa$
indexed by the positive real parameter $\kappa$.
These random curves are the only ones which
combine conformal invariance and a Markovian-type property (which is usually
already satisfied in the discrete setting). Provided
that the scaling limit of an interface in a model studied in
statistical physics (such as Ising, Potts or percolation)
exists and is conformally invariant (and this approach allows one
to give a precise meaning to this), then  the limiting object
must   be one of the
SLE$_\kappa$ curves.
Conformal invariance
 has now been rigorously shown  in some cases
(critical site percolation on the triangular
lattice has been solved by Stas Smirnov \cite {Sm},
the case of loop-erased random walks and uniform spanning
trees is treated in Lawler-Schramm-Werner \cite {LSWlesl}).
For a general discussion of the conjectured relation between the
discrete models and SLE, see \cite {RS}. See also \cite {LSWsaw}
for  self-avoiding walks and self-avoiding polygons.

In the SLE setting, the
critical exponents  simply correspond to  principal eigenvalues
 of some differential operators, see Lawler-Schramm-Werner
\cite {LSW1,LSW2,LSW3,LSW5}. Recognizing this led to  complete mathematical derivations of the values  of
critical exponents for the models, that have been proved to
be conformally invariant, in particular for critical percolation
on the triangular lattice (see \cite {SW}).
In order to establish  rigorously the conjectures for the other models,
the missing step is to show their
conformal invariance.

Using the Markovian property (which implies that with ``time''
the conditional probabilities of macroscopic  events are
martingales) of SLE and It\^o's formula, one readily sees
 that the probabilities of macroscopic events such as
crossing probabilities have to satisfy some second order
differential equations \cite {LSW1,LSW2,LSW3,Sper}.
This enables one to recover Cardy's formula in the case of SLE$_6$,
and to generalize it to other models (i.e. for
other values of $\kappa$).
Note that just as observed by Carleson in the case of
critical percolation, these crossing probabilities
formulae become extremely simple in well-chosen triangles,
 as pointed out by Dub\'edat \cite {Du}.

It is therefore  natural to think that SLE should be related to
conformal field theory and
to highest-weight representations of the Virasoro Algebra.
 Bauer-Bernard \cite {BB, BB2} recently
viewed (with a physics approach) SLE as a process living on
a ``Virasoro group'', which shows such a link
and enables them among other things to recover in conformal
field theory language, the generalized crossing probabilities mentioned
above.

Back in 1999, Lawler and Werner \cite {LW2} had introduced a notion of
universality based on a family of conformal restriction measures,
that gave a good insight into the fact that the exponents associated
to self-avoiding walks, critical percolation and simple random walks 
were in fact the same (these correspond in CFT language to the models 
with zero central charge) and pointed out the important role played
by these restriction properties (which became also instrumental in the 
papers \cite {LSW1,LSW2,LSW3}). 
In the recent paper \cite {LSWr} by Lawler, Schramm and Werner,
closely related (but slightly different) restriction properties are studied.  Loosely speaking
(and this will be recalled in more precise terms below), one looks
for random subsets $K$ of a given set (the upper half-plane, say),
joining two boundary points ($0$ and infinity, say), such that the
law of $K$ is invariant under the following operations:
For all simply connected subset $H$ of $\H$, the law of $K$
conditioned on $K \subset H$ is equal to the law of $\Phi (K)$,
where $\Phi$ is a conformal map from $\H$ onto $H$
preserving the two prescribed boundary points. In some sense, the 
law of $K$ is ``invariant'' under perturbation of the boundary.
It turns out that one can fully classify these random sets
(it is a one\--parameter family termed restriction measures, that are 
indexed by their positive real exponent),
and that they can be constructed in different but equivalent ways.
For instance,
by taking the hull of Brownian excursions (possibly reflected
on the boundary of the domain), or by adding to an SLE$_\kappa$
path a certain Poissonian cloud of Brownian loops. This gives an alternative description of the SLE
curves, that does not rely on Loewner's equation and on the Markovian property,
but can be interpreted as a variational equation (``how does the law of the 
SLE change'') with respect to perturbations of the domain. This in turn can be shown to correspond in the geometric
setting of CFT to differentiating the partition function with respect to the moduli, which then gives the correlation functions of
the stress\--energy tensor. In fact, the SLE correlation functions derived below, are those of the stress tensor. This will not be 
further explained in the present text, but is one of the subjects of the forthcoming paper \cite {FK}.

The aim of the present paper is to point out that these restriction
properties (and their relation to the SLE curves) can be rephrased
in a way that exhibits a direct and simple link between
the SLE curves (and therefore also the two-dimensional critical
systems) and representation theory. In this setting,
the Ward identities turn out to be  a reformulation of the
restriction property. More precisely, we will associate to each 
restriction measure a highest-weight representation of ${\cal A}$
(viewed as operators on a properly defined vector space).
The degeneracy of the representation corresponds to the Markovian
type property of SLE. The density of the Poissonian cloud of 
Brownian loops that one has to add to the SLE$_\kappa$ is (up to 
a sign-change) the central charge associated to the representation and
the exponent of the restriction measure is its highest-weight. 
 
The reader acquainted with
conformal field theory will recognize almost all the
identities that we will derive as ``usual and standard''
facts from the CFT perspective, but the point is here to give them
a rigorous meaning and interpretation in terms of SLE and discrete models.
Also, in the spirit of the conclusion of Cardy's review paper
\cite {Ca3} and as already confirmed by \cite {BB},
the rigorous SLE approach should hopefully  become useful and 
exploited within the theoretical physics community.

\section {Background}

\subsection {Chordal SLE}
The chordal SLE$_\kappa$ curve $\gamma$
is characterized as follows: 
The conformal maps
$g_t$ from $\H \setminus \gamma [0,t]$ onto $\H$ 
such that $g_t (z) = z + o(1) $ when $z \to \infty$
solve 
the ordinary differential equation 
$\partial_t g_t (z) =
2 /( g_t(z) - W_t)$ 
(and are started from $g_0(z) = z$),
where
$W_t = \sqrt {\kappa} b_t$ (here and in the 
sequel, $(b_t, t \ge 0)$ is a standard real-valued 
Brownian motion with $b_0=0$).
In other words, $\gamma_t$ is precisely the
point such that $g_t(\gamma_t) = W_t$.
See e.g. 
\cite {LSW1, RS} for the definition and 
properties of  SLE, or \cite {Lin, Wstf} for reviews.
Note that for any finite set of points, if one defines
the function 
$f_t (z) = g_t (z) - W_t$, the Markov property of the 
Brownian motion $b$ shows  that the 
law of $(f_{t_0 +t}, t \ge 0)$ is identical to that 
of $(f_t, t \ge 0)$. Then It\^o's formula immediately implies that 
for any set of real points $x_1, \ldots, x_n$ and any smooth
function $F: \R^n \to \R$,
\begin {eqnarray*}
\lefteqn{ dF (f_t (x_1) , \ldots, f_t (x_n))
= 
-dW_t \sum_{j=1}^n \partial_{j} F (f_t (x_1), \ldots ,f_t ( x_n))
}\\
&&+ 
dt 
\left\{ \frac {\kappa}{2}
(\sum_{j=1}^n \partial_j)^2 + (\sum_{j=1}^n \frac {2}{f_t (x_j)} 
\partial_j ) \right\} F (f_t (x_1), \ldots, f_t( x_n))
\end {eqnarray*}
i.e. if one defines the operators $L_N := - \sum_{j=1}^n x_j^{1+N} \partial_j$,
and the value  
$F_t = F(f_t(x_1), \ldots, f_t(x_n))$,
$$
dF_t = dW_t L_{-1}F_t + dt (\kappa/2 L_{-1}^2 - 2 L_{-2} ) F (f_t(x_1),
\ldots, f_t (x_n))
.$$
From this the chordal crossing probabilities \cite {LSW1,LSW3} are 
identified by using the fact that the drift term vanishes iff $F$
is a martingale i.e. if
$(\kappa/2 L_{-1}^2 - 2 L_{-2} )F=0$.
This already enabled  \cite {BB}  to tie a link 
with conformal field theory. 

\subsection {Chordal restriction}

All the facts recalled in this section are derived in \cite {LSWr}.
Let $\H$ denote the open upper half-plane.
We call ${\cal H}_+$ (resp. ${\cal H}$)
 the family of simply connected subsets $H$ of $\H$ such that:
$\H \setminus H$ is bounded and bounded away from $\R_-$
(resp. from $0$).
For such an $H$, we define the conformal map $\Phi_H$ from
$H$ onto $\H$ such that $\Phi_H (0) = 0$ and $\Phi_H (z) \sim z$
when $z \to \infty$.

We say that a simply connected set
$K$ in $\H$ satisfies the
``one-sided restriction property'' (resp. the two-sided
restriction property) if:

\begin {itemize}
\item It is scale-invariant (the laws of $K$ and of $\lambda K$
are identical for all $\lambda>0$).
\item
For all $H \in {\cal H}_+$ (resp. $H \in {\cal H}$),
 the conditional law of $\Phi_H (K)$
given $K \cap (\H \setminus H)  = \emptyset$ is
identical to the law of $K$.
\end {itemize}

All such random sets $K$ are classified in \cite {LSWr}.
It is not difficult to see that
this definition implies that, for all $H \in {\cal H}_+$
(resp. $H \in {\cal H}$), and for  some fixed exponent
$h >0$,
$$
P [ K \cap ( \H \setminus H) = \emptyset ] = \Phi_H'(0)^h
.$$
This (modulo filling) in fact characterizes the law of the random
set $K$.
Conversely, for all $h >0$, there exists
such a random set $K$.
It can be constructed through  three a priori very different means:
By using a variant of SLE$_{8/3}$, called SLE$(8/3, \rho)$,
by filling certain (reflected) Brownian excursions (see below), or by adding
Brownian loops to a certain SLE$_\kappa$.
In the two-sided case, such random sets $K$ only exist when $h \ge 5/8$.
The only value $h$ corresponding to a simple curve $K$ is
$h=5/8$ (and this random curve conjecturally corresponds
to the scaling limit of half-plane infinite self-avoiding walks,
see \cite {LSWsaw}).

\begin{figure}
\centerline{\includegraphics*[height=1.5in]{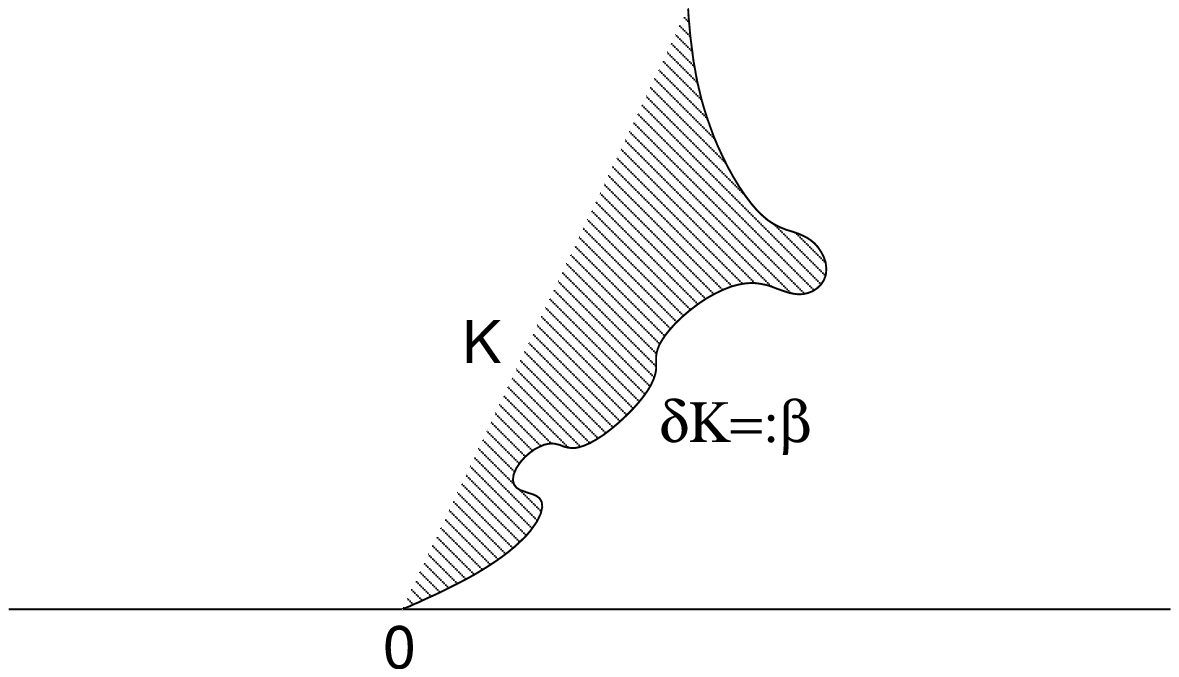}}
\caption{\label{f.1}The set $K$ and its right-boundary $\beta$.}
\end{figure}

Here we will focus mainly  on the right boundary of such
sets $K$ (which -in the one-sided case- is
an equivalent way of describing $K$) that
will be denoted by $\beta$. It is shown in \cite {LSWr} that
this curve is an SLE$(8/3, \rho)$ for some $\rho=\rho (h)$.
In particular, the Hausdorff dimension of all these curves $\beta$
is $4/3$.

The most important examples of such sets $\beta$ are:
\begin {itemize}
\item
The SLE$_{8/3}$ curve itself. In fact, it is the only simple curve satisfying the 
two-sided restriction property. The corresponding exponent $h$ is $5/8$.
\item
If one takes the ``right-boundary'' of a Brownian excursion from 0 to $\infty$ in the 
upper-half plane (this process is a Markov process that can be loosely 
described as Brownian motion conditioned   never to  hit the real line). 
This corresponds to the exponent $h=1$.
\end {itemize}

This last example can in fact be generalized to all $h<1$: If one takes the 
``right-boundary'' of a Brownian motion started from the origin that is 
\begin {itemize}
\item
Conditioned never to hit the positive half-axis,
\item
Reflected off the negative half-axis with a fixed well-chosen angle $\theta(h)$,
\end {itemize}
then, it satisfies the one-sided restriction property with exponent $h$.
See \cite {LSWr} for more details.

Also, it is easy to see that
 if $\beta_1, \beta_2, \ldots, \beta_N$ are $N$ such independent curves with respective exponents
$h_1, h_2, \ldots,  h_N$,
 then the right-boundary $\beta$ of $\beta_1 \cup \ldots \cup \beta_N$ also satisfies the 
one-sided restriction property with exponent $h_1 + \cdots + h_N$.
This is simply due to the fact that 
$$
P [ \beta \cap (\H \setminus H) = \emptyset ]
= 
\prod_{j=1}^{j=N} P [ \beta_j \cap (\H \setminus H) = \emptyset ]
=
\Phi_H' (0)^{h_1 +  \cdots + h_N}
$$
for all $H \in {\cal H}_+$.

In particular, this shows that any one-sided restriction measure can be constructed using 
the union of independent (conditioned and reflected) Brownian motions.

\section {Boundary correlation functions}

Suppose now that the random 
simple curve $\beta$ satisfies the one-sided restriction property.
For each real positive $x$ and $\eps$,
define the event
$$
E_\eps (x):= \{ \beta \cap [x, x + i \eps \sqrt {2} ] \not= \emptyset \}.
$$
The one-sided restriction property of $\beta$ shows that
$$P [ E_{\eps_1} (x_1) \cup \ldots \cup E_{\eps_n} (x_n)]
= 1 - \Phi_{\H \setminus \cup_{j=1}^n [x_j, x_j + i \eps_j \sqrt {2}]}'(0)^h
,$$
for all positive $x_j$'s and $\eps_j$'s. 
These derivatives can (in principle) be determined ($\Phi^{-1}$ is a simple 
Schwarz-Christoffel transformation, see \cite {Ah}).
This (by a simple inclusion-exclusion
formula) yields the values of the probabilities
$$
f(x_1, \eps_1, \ldots, x_n , \eps_n)
:=P [ E_{\eps_1}(x_1) \cap \ldots \cap E_{\eps_n} (x_n) ]
$$
in terms of $x_1, \ldots, x_n, \eps_1, \ldots, \eps_n$.
For example, when $n=1$,
$$ 
f(x, \eps) = P [ E_\eps (x) ] = 1 - \left( \frac {x}{\sqrt { x^2 + 2 \eps^2}} \right)^h
.$$
In particular, $f(x, \eps) \sim \eps^2 h / x^2$ when $\eps \to 0$. We then define 
$B_1^{(h)} (x)= h/x^2 = \lim_{\eps \to 0} \eps^{-2} f(x, \eps)$. 

More generally, one can define the functions $B_n = B_n^{(h)}$
as
\begin {equation}
\label {Bndef}
B_n ( x_1, \ldots, x_n)
:=
\lim_{\eps_1 , \ldots, \eps_n \to 0}
\eps_1^{-2} \ldots \eps_n^{-2}
f(x_1, \eps_1, \ldots, x_n, \eps_n )
.\end {equation}

\begin{figure}
\centerline{\includegraphics*[height=1.5in]{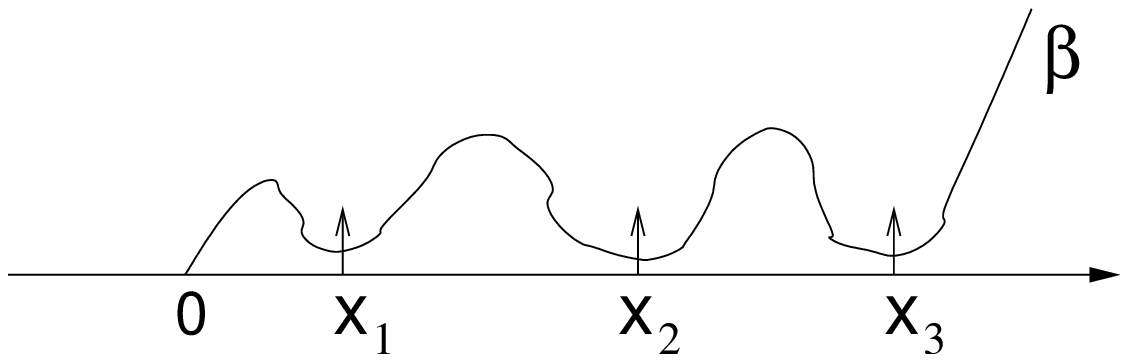}}
\caption{\label{f.2}The event $E= E_\eps (x_1) \cap E_\eps (x_2) \cap E_\eps (x_3)$.}
\end{figure}

An indirect way to justify the existence of the existence of the limit in (\ref {Bndef})
goes as follows:
First, note that when $h=1$, the description of
$\beta$ as the right-boundary of a Brownian excursion 
 yields  the existence of $B_n^{(1)}$ and the 
 following explicit expression:
$$
B_n^{(1)}
(x_1, \ldots, x_n )
= \sum_{s \in \sigma_n}
\prod_{j=1}^{n-1}
(x_{s(j)} - x_{s(j-1)})^{-2}
,$$
where $\sigma_n$ denotes the group of permutations of $\{1, \ldots, n\}$
and by convention $x_{s(0)} = 0$. This is due to the fact that $\beta$
intersects all these slits if and only if the Brownian excursion itself
intersects all these 
slits. One then decomposes this event according to the order with which the excursion
actually hits them, and one uses its strong Markov property.

Similarly, an analogous reasoning using the Brownian motions reflected on the 
negative half-axis, and conditioned not to hit the positive half-axis (and its strong Markov property), yields the existence of the limit in (\ref {Bndef}) for all $h<1$. 

Also, since the right-boundary of the union $K_1 \cup \ldots \cup K_N$ of $N$ independent 
sets satisfying the restriction property with
exponents $h_1, \ldots, h_N$ satisfies the 
one-sided restriction property with exponent $h_1+ \cdots  + h_N$, we get easily 
the existence of the limit in (\ref {Bndef}) for all $h$ (using the existence 
when $h_1, \ldots, h_N \le 1$), and 
the following property of the functions $B$:
For all $R: \{1, \ldots, n \} \to \{1, \ldots, N\}$,
write
$r(j) = \hbox {card} ( R^{-1} \{ j\})$. Then,
\begin {equation}
\label {sg}
B_n^{(h_1  + \cdots + h_N)} (x_1, \ldots, x_n)
= \sum_R \prod_{j=1}^N B_{r(j)}^{(h_j)} ( x_{R^{-1} \{ j \} })
,\end {equation}
where $B_0 = 1$ and $x_I$ denotes the vector with coordinates
$x_k$ for $k \in I$.
This yields a simple explicit formula for $B^{(n)}$ when $n$ is 
a positive integer.

In the general case, one way to compute $B_n^{(h)}$
is to use the following inductive relation (together with the 
convention $B^{(h)}_0 \equiv 1$):

\begin {proposition}
For all $n \in \N$, $x, x_1, \ldots, x_n \in \R_+$,
\begin {eqnarray}
\nonumber
\lefteqn{ B_{n+1}^{(h)} ( x, x_1, x_2, \ldots, x_n )
\ = \  \frac {h}{x^2} B_n^{(h)} (x_1, \ldots , x_n)} \\
&&
-\sum_{j=1}^n \left\{ (\frac {1}{x_j-x} + \frac 1x ) \partial_{x_j}
  - \frac {2}{(x_j-x)^2} \right\} %\right]
 B_n^{(h)} (x_1, \ldots , x_n).
\label {ward1}
\end {eqnarray}
\end {proposition}
This relation plays the role of the Ward identities in the CFT
formalism.

\medbreak
\noindent
{\bf Proof.}
Suppose now that the real numbers $x_1, \ldots, x_n$ are fixed and
let us focus on the event 
$E=E_\eps (x_1) \cap \ldots \cap E_\eps (x_n)$.
Let us also choose another point $x \in \R$ and a small $\delta$.
Now, either the curve $\beta$
avoids $[x , x+ i \delta \sqrt {2}]$ or it does hit it.
This additional slit is hit (as well as the $n$ other ones)
with a probability $A$ comparable
to
$$\eps^{2n} \delta^2 B_{n+1} (x_1, \ldots, x_n, x)$$
when both $\delta$ and $\eps$ vanish.
On the other hand, 
the image 
of $\beta$ conditioned to avoid $[x, x+i \delta \sqrt {2}]$ 
under the map
$$\varphi 
(z) = \Phi_{\H \setminus [x, x+ i \delta \sqrt {2}]}
= \sqrt { (z-x)^2 + 2 \delta^2 } - \sqrt { x^2 + 2 \delta^2}
$$
 has the same law as $\beta$.
In particular, we get immediately that 
\begin {eqnarray*}
A' &:= &
\Prob [  E
\mid \beta \cap [x, x + i \delta \sqrt {2} ] = \emptyset ]
 \\
&\sim&
\eps^{2n} \prod_{j=1}^n |\varphi'(x_j)|^2
B ( \varphi(x_1), \ldots, \varphi (x_n) )
\end {eqnarray*}
when $\eps \to 0$ (this square for the derivatives can be interpreted as 
 the fact that the
``boundary exponent'' for restriction measures is always 2). 
But when $\delta$ vanishes, 
$$
\varphi (z) = z +  {\delta^2} \left(
\frac 1 {z-x} + \frac 1 x \right) + o (\delta^2)
$$
and
$$
\varphi'(z) = 1 - \frac {\delta^2}{(z-x)^2} + o (\delta^2).
$$
On the other hand,
\begin {equation}
\label {e.exp}
\Prob [ E ] 
= 
A + A' \Prob [ \beta \cap [x, x+i \delta \sqrt {2} ] = \emptyset ]
\end {equation}
is independent of $\delta$
and
$$
\Prob [ \beta \cap [x, x+ i \delta \sqrt {2} ] = \emptyset ]
= \varphi'(0)^{h}
= 1 - \frac {h \delta^2}{x^2} + o (\delta^2)
$$
when $\delta \to 0$.
Looking at the $\delta^2$ term in the $\delta$-expansion of
(\ref {e.exp}), we get (\ref {ward1}). 
\qed

\section {Highest-weight representations}

We now define, for all $N \in \Z$,
the operators
$$
{\cal L}_N = \sum_j
\{ - x_j^{1+N} \partial_{x_j} - 2 (N+1) x_j^N \}
$$
acting on functions of the
real variables $x_1, x_2 , \ldots $.
In fact, one should  (but we will omit this) make precise the range of $j$ i.e.
define ${\cal L}_N$ on the union over $n$ of the spaces $V_n$ of functions
of $n$ variables $x_1, \ldots, x_n$. 

Note that these operators satisfy the commutation
relation
$$[ {\cal L}_N, {\cal L}_M ]
= (N-M) {\cal L}_{N+M}
$$
just as the operators $L_N$ do.
In other words, the vector space generated by these operators is
(isomorphic to)
the the Witt algebra, i.e. the Lie algebra of vector fields on the unit circle (this is 
classical, see e.g. \cite {FF}).

Note also that one can rewrite the Ward identity in terms of these
operators as:
\begin {equation}
\label {ward2}
B_{n+1}^{(h)} (x, x_1, \ldots, x_n)
= \frac {h}{x^2} B_n^{(h)} (x_1, \ldots, x_n)
+ \sum_{N \ge 1} x^{N-2} {\cal L}_{-N} B_n^{(h)}  (x_1, \ldots, x_n).
\end {equation}
We are now going to consider vectors $w = (w_0, w_1, w_2, \ldots)$
such that for each $n$, $w_n$ is a function of
$n$ variables $x_1, \ldots, x_n$.
An example of such a vector is 
$$B=B^{(h)}=
(B_0^{(h)}, B_1^{(h)}, B_2^{(h)}, \ldots )$$
where $B_0^{(h)}$ is set to be equal to $1$. For convenience we will fix $h$ and not always 
write the $(h)$ superscript.

For such a vector $w$, we define for all $N \in \Z$ the
operator $l_N$
in such a way that
$$
w_{n+1}
(x, x_1, \ldots, x_n)
= \sum_{N \in \Z} x^{N-2} ( l_{-N}(w))_n (x_1, \ldots, x_n)
.$$
In other words, the $n$-variable component $(l_N (w))_n$ of $l_N(w)$
is the $x^{-N-2}$ term in the Laurent expansion of $w_{n+1} (x, x_1, \ldots, x_n)$
with respect to $x$.

For example, the Ward identity (\ref {ward2}) gives the values of $l_N (B)$: 
\begin {equation}
\label {hwv}
l_N (B) = 
\left\{ 
 \begin {array}{l@{\hbox { if }}l}
 (0, 0, \ldots ) & N>0 \\
(h B_0, h B_1, \ldots )  & N=0 \\
({\cal L}_N B_0, {\cal L}_N B_1, \ldots)  & N<0
\end {array}
\right.
\end {equation}
We insist on the fact that $l_N (B)$ does not coincide with 
${\cal L}_N (B)$ for non-negative $N$'s. For instance,
$$
{\cal L}_0 ( B_1) = 0 \not= h B_1 = (l_0 B)_1.
$$
But the identity for negative $N$'s can be iterated as follows:
\begin {lemma}
\label {llemma}
For all $k \ge 1$ and  negative $N_1, \ldots, N_k$,
\begin {equation}
\label {elemma}
 ( l_{N_1} \cdots l_{N_k} B )_n
= {\cal L}_{N_1} \ldots {\cal L}_{N_k} B_n.
\end {equation}
\end {lemma}
\medbreak
\noindent
{\bf Proof of the Lemma.}
This is a rather straightforward consequence of (\ref {ward2}).
We have just seen that it holds for $k = 1$.
Assume that (\ref {elemma}) holds for some given integer $k \ge 1$.
Then, for all negative $N_2, \ldots, N_k$,
\begin {eqnarray*}
\lefteqn{
({\cal L}_{N_2} \cdots {\cal L}_{N_k} B)_{n+1}
(x, x_1, \ldots, x_n)
}\\
&=&
u + \sum_{N \le -1} x^{-N-2} {\cal L}_{N} {\cal L}_{N_2} \ldots 
{\cal L}_{N_k} B_n (x_1, \ldots, x_n)
\end {eqnarray*}
where $u$ is a Laurent series in $x$
such that $u(x, x_1, \ldots, x_n) = O (x^{-2})$
when $x \to \infty$.
We then apply ${\cal L}_{N_1}$ (viewed as acting on the space of 
functions of the $n+1$ variables $x,x_1, \ldots, x_n$)
 to this equation, where $N_1<0$.
There are two $x^{-N-2}$ terms in the expansion on the right-hand side:
The first one is simply
$$
x^{-N-2}
{\cal L}_{N_1} {\cal L}_{N} {\cal L}_{N_2} \ldots 
{\cal L}_{N_k} B_n (x_1, \ldots, x_n)
.$$
The second one comes from the term 
\begin {eqnarray*}
\lefteqn{
({\cal L}_{N_1} x^{-N-N_1-2}) {\cal L}_{N+N_1} {\cal L}_{N_2} 
\ldots {\cal L}_{N_k} B_n (x_1, \ldots, x_n)
} \\
&=&
 (N-N_1) x^{-N-2} {\cal L}_{N+N_1} {\cal L}_{N_2} 
\ldots {\cal L}_{N_k} B_n (x_1, \ldots, x_n).
\end {eqnarray*}
The sum of these two contributions is indeed
$$
x^{-N-2}
{\cal L}_N {\cal L}_{N_1} \ldots {\cal L}_{N_k} B_n (x_1, \ldots, x_n)$$
because of the commutation relation 
$$ {\cal L}_{N_1} {\cal L}_N + (N-N_1) {\cal L}_{N+N_1}
= {\cal L}_N {\cal L}_{N_1}
.$$
This proves (\ref {elemma}) for $k+1$.
\qed

\medbreak
We now define, the vector space $V$
generated by the vector $B$ and all vectors
$l_{N_1} \ldots l_{N_k} B$
for negative
$N_1, \ldots, N_k$ and positive $k$
(we will refer to  these vectors 
as the generating vectors of $V$).
Then:
\begin {proposition} For all $v \in V$, for all $M,R$ in $\Z$,
$$l_M (v) \in V \hbox { and }
[ l_M , l_R ] v = (M-R) l_{M+R} v
.$$
\end {proposition}

We insist again on the fact that $l_N$ only coincides with ${\cal L}_N$
for negative $N$. Also, the commutation relation for the $l_N$'s
does not hold for a general vector. The above statement only
says that it is valid on this special vector space $V$.

\medbreak
\noindent
{\bf Proof.}
Note that the commutation relation holds
for negative $R$ and $M$'s because
of Lemma \ref {llemma}.

Suppose now that 
$N_1, \ldots, N_k$ are negative.
Then,
\begin {eqnarray*}
{\cal L}_{N_1} \ldots {\cal L}_{N_k}
B_{n+1}
&=&
\sum_{N \le 0, I}
{\cal L}_{N_{i_1}} \ldots {\cal L}_{N_{i_r}} (x^{-2-N}) \\
&& \hskip 1cm
\times
{\cal L}_{N_{j_1}} \ldots {\cal L}_{N_{j_s}} (l_N B)_n (x_1, \ldots, x_n)
\end {eqnarray*}
where the sum is over all $I:=\{ i_1 , \ldots, i_r \}
\subset \{1, \ldots, k \}$. One then writes
$\{ j_1, \ldots j_s\} = \{1, \ldots, k \} \setminus 
\{ i_1, \ldots, i_r \}$ (and the $i$'s and $j$'s are increasing).
We use $l_N (B)_n$ instead of ${\cal L}_N B_n$
 to simplify the expression (otherwise
the case $N=0$ would have to be treated separately).

Since 
\begin {eqnarray*}
\lefteqn{
{\cal L}_{N_{i_1}} \ldots {\cal L}_{N_{i_k}}
(x^{-2-N})
}\\
&=& 
(N - 2 N_{i_r}) (N - N_{i_r} -
2N_{i_{r-1}}) \ldots \\
&& \ldots (N - N_{i_r} - \ldots 
-N_{i_2} - 2 N_{i_1}) x^{-2-N+N_{i_1} + \cdots + N_{i_k}},
\end {eqnarray*}
it follows immediately  that for all integer $M$,
\begin {eqnarray}
\nonumber
\lefteqn {(l_M l_{N_1} \ldots l_{N_k} B)_n
}\\
&=&
\sum_{I : \ M+N_{i_1} + \cdots +N_{i_r} \le 0}
(M+N_{i_1} + \ldots + N_{i_{r-1}} - N_{i_r}) \ldots (M - N_{i_1}) 
\nonumber \\
&& \hskip 1cm
\times 
 {\cal L}_{N_{j_1}} \ldots {\cal L}_{N_{j_s}} (l_{M+N_{i_1} + \ldots + N_{i_r}}
  B)_n
\label {compl}
\end {eqnarray}
This implies that indeed, $l_M (V) \subset V$.
When $M \le 0$, then for any $i_1, \ldots, {i_r}$, $M +N_{i_1}
+ \ldots + N_{i_r} \le 0$, so that 
the sum is over all $I$.

Suppose now that $M \ge 0$, $R < 0$, and consider 
$v = l_{N_1} \ldots l_{N_k}$ for some fixed negative
$N_1, \ldots, N_k$.
We can apply (\ref {compl}) to get the expression of 
$l_{R+M} v$, of $l_M l_R v$ and of $l_M v$.
Furthermore, we can use the Lemma to deduce the following
expression for $l_R l_M v$:
\begin {eqnarray*}
\lefteqn {(l_R l_M v)_n}\\
&=&
\sum_{I : \ M+N_{i_1} + \cdots +N_{i_r} \le 0}
(M+N_{i_1} + \ldots + N_{i_{r-1}} - N_{i_r}) \ldots (M - N_{i_1}) \\
&& \hskip 1cm
\times 
 {\cal L}_R{\cal L}_{N_{j_1}} \ldots {\cal L}_{N_{j_s}} (l_{M+N_{i_1} + \ldots + N_{i_r}}
  B)_n
  \end {eqnarray*}
On the other hand,
\begin {eqnarray*}
\lefteqn{
(l_M l_R v)_n
}
\\
&=&
\sum_{I_0 : \ M+N_{i_0} + \cdots +N_{i_r} \le 0}
(M+N_{i_0} + \ldots + N_{i_{r-1}} - N_{i_r}) \ldots (M - N_{i_0}) \\
&& \hskip 1cm
\times 
 {\cal L}_{N_{j_1}} \ldots {\cal L}_{N_{j_s}} (l_{M+N_{i_0} + \ldots + N_{i_r}}
  B)_n,
  \end {eqnarray*}
where this time, the sum is over 
$\{ i_0, \ldots, i_r \} \subset \{ 0, \ldots, k \}$, and
we put $R= N_0$.
The difference between these two expressions is due to the 
terms (in the latter) where $i_0 =0$:
\begin {eqnarray*}
\lefteqn { [ l_M, l_R] v } \\
&=&
(M-R) 
\sum_{I : \ M+N_{i_1} + \cdots +N_{i_r} \le 0}
(M+R+N_{i_1} + \ldots + N_{i_{r-1}} - N_{i_r}) \ldots\\
&& \ldots  (M+R - N_{i_1}) 
 {\cal L}_{N_{j_1}} \ldots {\cal L}_{N_{j_s}} (l_{M+R+N_{i_1} + \ldots + N_{i_r}}
  B)_n\\
  &=& 
  (M-R) l_{M+R}.
\end {eqnarray*}
This proves the commutation relation for negative $R$
and arbitrary $M$.

Finally, to prove the commutation relation when both $R$ and 
$M$ are negative and
$v = l_{N_1} \ldots l_{N_k}$ 
as before, it suffices to use the previously proved
commutation relations to write 
$l_M v$, $l_R v$ and $l_{M+R} v$
as linear combination of the generating vectors of $V$.
Then, one can iterate this procedure to 
express $[l_M, l_R] v$ as a linear combination of 
the generating vectors of $V$. Since this formal algebraic calculation
is identical to that one would do in the Lie algebra ${\cal A}$, 
one gets  indeed 
$[l_M, l_R] v = (M-R) l_{M+R}$, 
which therefore also holds for any $v \in V$. 
\qed

\medbreak

To put it differently,  to each
(one-sided) restriction measure, one can simply
associate a highest-weight representation of the Lie algebra
${\cal A}$ (without central extension) acting on a
certain space of function-valued vectors.
The value of the highest weight is the exponent 
of the restriction measure.

Note that the right-sided boundary of a simply connected
set $K$ satisfying the two-sided  restriction
property satisfies the one-sided restriction property
(so that one can also
associate a representation to it).
In this case,
the function $B_n$ also represents the limiting value
of
$$
\eps^{-2n} 
P ( K \hbox { intersects all slits } [x_j, x_j + i \eps \sqrt {2}] ,
j =1 , \ldots , n )
$$
even for negative values of some $x_j$'s.

\section {Evolution and degeneracy}
\subsection {SLE$_{8/3}$}

We are now going to see how to combine the previous 
considerations with a Markovian property. For instance, 
does there exist a value of $\kappa$ such that SLE$_\kappa$
satisfies the restriction property? We know from \cite {LSWr}
that the answer is yes, that the value of $\kappa$ is $8/3$ 
and that the corresponding exponent is $5/8$. 
This  
``boundary exponent'' for SLE$_{8/3}$ has appeared before in the 
theoretical physics literature (see e.g. \cite {DS})
as the boundary exponent for 
long self-avoiding walks (which is consistent with the conjecture 
\cite {LSWsaw} that this SLE is the scaling limit of the 
half-plane self-avoiding walk). 
This exponent was identified as the only
possible highest-weight of a highest-weight representation 
of ${\cal A}$ that is {\em degenerate} at level two.

We are now going to see that indeed, the Markovian property of SLE
is just a way of saying that the two vectors
$l_{-2} (B)$ and $l_{-1}^2 (B)$
 are not 
independent. This shows (without using the computations in \cite {LSWr})
why the values $\kappa=8/3$, $h=5/8$ pop out.

Suppose that $\beta$ is an SLE$_\kappa$.
Consider the event 
$E:=
E_{\eps_1} ( x_1) \cap \ldots \cap E_{\eps_n} (x_n)$ 
as in the definition of $B_n^{(h)}$.
If one considers the 
conditional probability of $E$ given $\beta$ up to time $t$, 
then it is the probability that an (independent) SLE 
$\tilde \beta$ hits the (curved) slits 
$f_t ([x_j, x_j + i \eps_j \sqrt {2}])$.
At first order, this is equivalent to hitting 
the straight slits
$$[f_t(x_j), f_t(x_j) + i \eps_j \sqrt {2} f_t'(x_j)].$$

If the SLE satisfies the restriction property
with exponent $h$, then this means that 
$$
f_t'(x_1)^{-2} \ldots f_t'(x_n)^{-2} B_n^{(h)} 
( f_t (x_1), \ldots, f_t (x_n) )
$$
is a local martingale. 
Recall that
$$
\partial_t f_t (x) =- \sqrt {\kappa} db_t + \frac {2 }{f_t (x)}
\hbox { and }
\partial_t f_t' (x) = \frac {-2 f_t'(x)}{f_t(x)^2}.
$$
Hence, since the drift term of the previous local martingale
vanishes,
It\^o's formula yields
$$
\frac {\kappa}{2} {\cal L}_{-1}^2 B_n - 2 {\cal L}_{-2} B_n 
= 0 
$$
for all $n \ge 1$.
Note that the operators are 
 ${\cal L}$'s, and not $L$'s as in the 
 crossing probability formulae,
 because of the local scaling properties 
 of the functions $B$.
 
 In other words, $l_{-2}(B)$ and $l_{-1}^2(B)$
 are collinear and the previously described highest-weight representation of 
 ${\cal A}$ must be degenerate at level two.
 It is elementary to deduce the values of $h$ and $\kappa$,
 using the fact that
 $$
 l_2 (\frac {\kappa}{2} l_{-1}^2 - 2 l_{-2})B
 = (3 \kappa - 8) l_0 B = 0
 $$
 which implies that $\kappa=8/3$
 and
 $$
 l_1 (\frac {\kappa}{2} l_{-1}^2 - 2 l_{-2} ) B
 = \frac {\kappa}{2} (4 l_{-1} l_0 B + 2l_{-1} B)
 -6 l_{-1} B =
 (2 \kappa h + \kappa - 6 ) l_{-1} B =0
 $$
 which then implies that $h = 5/8$.
 
\subsection {The cloud of bubbles}
 
We are now going to use the description of the 
``restriction paths'' $\beta$ via SLE curves to which one adds a Poissonian cloud of
Brownian bubbles, as explained in \cite {LSWr}. Let us briefly recall how it goes. 
Consider an SLE$_\kappa$ for $\kappa< 8/3$. As we have just seen, it does not 
satisfy the restriction property. However, if one adds to this curve 
an appropriate random cloud of Brownian loops, then the obtained set 
satisfies the two-sided  restriction property for a certain exponent
$h > 5/8$ (and its right-boundary $\beta$ satisfies the one-sided 
restriction property). More details and properties of the Brownian 
loop-soup and the procedure of adding loops can be found in \cite {LSWr,LW}.

Intuitively  this phenomenon can be understood from  the case, where $\kappa=2$:
SLE$_2$ is the scaling limit of the loop-erased 
random walk excursion (see \cite {LSWlesl}). Adding Brownian loops to 
it, one should (in principle) recover the Brownian excursion that 
satisfies the restriction property with parameter $h =1$.

More generally, let $\kappa < 8/3$ be fixed, and consider an SLE$_\kappa$
curve $\gamma$, with its usual time-parametrization. 
There exists a natural (infinite) measure on Brownian bubbles in $\H$ rooted at the origin.
This is a measure supported on Brownian paths of finite length 
in $\H$ that start and end at the origin (more generally, we say that a bubble in $H$ rooted
at $x \in \partial H$ is a path $\eta$ of finite length $T$ such that 
$\eta (0,T) \in H$ and $\eta (0)= \eta (T) = x$).
Consider a Poisson point process of these Brownian bubbles in $\H$, with intensity
$\lambda$ (more precisely, $\lambda$ times the measure on Brownian bubbles). A realization of this point process is a family $(\hat \eta_t, t \ge 0)$  
such that for all but a random countable set $\{t_j\}$ of times, $\hat \eta_t = \emptyset$ and
for the times $t_j$, $\hat \eta_{t_j}$ is a (Brownian) bubble in $\H$ rooted at the origin.
We then define for all $t$, $\eta_t = f_t^{-1} (\hat \eta_t)$, so that $\eta_t$ is empty
if $t \notin \{t_j\}$ and is a bubble in $\H \setminus \gamma [0, t_j]$ rooted at
$\gamma (t_j)$ if $t = t_j$.
Another equivalent way to define this random family $(\eta_t, t\ge 0)$ via a 
certain Brownian loop-soup is described in \cite {LW}.

Define the union $\Gamma$ of
 $\gamma$ and the bubbles $\eta_t$, i.e.
 $$ 
 \Gamma = \cup_{t \ge 0} (\{ \gamma_t \} \cup \eta_t).
 $$
We let ${\cal F}_t$ denote the $\sigma$-field generated by $(\gamma_s, \eta_s, s \le t)$.
 
The right outer-boundary $\beta$ (see \cite {LSWr,LW}) of 
$\Gamma$ then satisfies the restriction property (actually $\Gamma$ satisfies the
two-sided restriction property).
This is proved in \cite {LSWr} studying the conditional probabilities 
that $\Gamma$ avoids a given set $A$ with respect to the filtration 
generated by $\gamma$ alone. 
As observed in \cite {LSWr}, the relation between the density $\lambda (\kappa)$ of the
loops that one has to add to the SLE$_\kappa$ and 
the exponent $h(\kappa)$ of the corresponding restriction measure
(i.e. $h= (6 - \kappa)/ 2\kappa$ and $\lambda = (8- 3 \kappa) h$) recalls the relation between the central charge and the 
highest-weight of degenerate highest-weight representations of the Virasoro algebra (which is the central extension of ${\cal A}$). We shall try in this subsection to give one
way to explain the relation to representations, via the functions $B_n^{(h)}$, and therefore recover these values of $h$ and $\lambda$, just assuming that if one adds the cloud of bubbles with intensity some $\lambda$, one obtains a restriction measure.

It is worthwhile emphasizing that in this context, the functions $B_n^{(h)}$ are only indirectly related to the SLE curve via this Poissonian cloud of loops. They do for instance not represent the probabilities that the SLE itself does visit the infinitesimal slits, but the probability that some loops that have been attached to this SLE curve do visits the 
infinitesimal slits. 

Recall that the functions $B_n^{(h)}$ are related to a highest-weight representation of 
${\cal A}$, as discussed in the previous section. As in the $\kappa=8/3$ 
case, we will try to obtain an additional information on this representation, using 
the evolution of the SLE curve. More precisely:
How does the (conditional) probability with respect to ${\cal F}_t$ 
of the event
$E$ that $\beta$ intersects the $n$ slits $[x_j, x_j + i \eps_j \sqrt {2}]$ 
for infinitesimal $\eps_j$'s evolve with time? Here is a heuristic discussion,
that can easily be made rigorous:

Consider an infinitesimal time $\Delta$. 
Let $\tilde \Gamma_\Delta$ denote the union of $\gamma [\Delta, \infty)$ and
the loops that it does intersect. More precisely, 
$$\tilde \Gamma_\Delta = \cup_{t> \Delta}(\{ \gamma_t \} \cup \eta_t).$$
Typically (for very small $\Delta$), there is no bubble 
$\eta_t$ for $t \in [0,\Delta]$ that does intersect one of these $n$ slits.
In this case, the conditional probability of the event $E$
given ${\cal F}_\Delta$
 is simply the probability  that $\tilde \Gamma_\Delta$ does intersect these $n$ 
slits (given ${\cal F}_\Delta$). The definition 
of $\gamma$ and of the bubbles show that the conditional law of 
$f_\Delta ( \tilde \Gamma_\Delta ) $ 
given ${\cal F}_\Delta$
is independent of $\Delta$ (in particular, it is the same as for $\Delta= 0$ i.e. the law
of $\Gamma$).
This shows that (exactly as in the $\kappa= 8/3$ case),
the conditional probability of $E$ has a drift term 
due to the distortion of space induced by the SLE (i.e. by $f_\Delta$) of the type
$$(\frac {\kappa}{2} {\cal L}_{-1}^2 B_n - 2 {\cal L}_{-2} B_n) \Delta.$$

But there is an additional term due to the fact that
one might in the small time-interval $[0, \Delta]$, have added a
Brownian loop $\eta_t$ to the curve that precisely goes through one or several of
the $n$ slits $[x_j, x_j +  i \eps_j \sqrt {2}]$.
The probability that one has added a loop that 
goes through the $j$-th slit is 
of order 
$ {\lambda  \eps_j^2} \Delta / {x_j^4}$.
This fact is due to  scale-invariance. Here $\lambda$ is
 the (constant) density of loops 
that is added on top of the 
SLE curve (we use this definition for
this density $\lambda$ in this paper, as in \cite {LSWr}; in 
other contexts, replacing $\lambda$ by $\lambda/6$ can be more natural). One way to 
understand the $\eps_j^2 /x_j^4$ term is that the Brownian bubble has to 
go from $0$ to the slit, which contributes a factor $\eps_j^2/x_j^2$, and then 
back to the origin, which contributes also $1/x_j^2$.
If such a loop has been added, the conditional probability of $E$ is 
(at first order) the probability that the SLE+loops hits the remaining
$n-1$ slits, i.e. 
$f_{n-1} (x_{\{ 1, \ldots n\} \setminus \{ j\}}) \prod_{l \not= j} \eps_l^2$
(here and in the sequel $x_J$ stands for $(x_{j_1}, \ldots, x_{j_p})$ when 
$J= \{ j_1, \ldots, j_p\}$).
More generally, define 
 $T_0 =0$, $T_1 (x)= 1/ x^4$, and for $p \ge 2$,
$$
T_p(x_1, \ldots, x_p)
= 
\sum_{s \in \sigma_{p}} 
\frac {1} {x_{s(1)}^2 (x_{s (2)} - x_{s (1)})^2
\ldots ( x_{s (p)} - x_{s (p-1)})^2 x_{s (p)}^2 }
.$$
Each $s$ corresponds intuitively to an order of visits of the infinitesimal 
slits by the loop.
For $J = \{ j_1, \ldots, j_{p} \}  \subset \{1, \ldots,n \} $
with $|J|=p \ge 1$, the probability to add a loop that goes precisely through the 
slits near $x_j$ for $j \in J$ is of the order of
$$
\eps_{j_1}^2 \ldots \eps_{j_{p}}^2 T_{p} ( x_J ) \lambda \Delta
.$$
We are therefore naturally led to define the operator $U$ by
$$
(Uf)_n (x_1, \ldots, x_n )
= \sum_{J \subset \{1 , \ldots, n \}}
T_{p} ( x_J ) \times  f_{n-p} (x_{\{ 1, \ldots n\} \setminus J }).
$$
Then, the fact that $P (E | {\cal F}_t ) $ is a martingale, 
shows that the drift term vanishes i.e. that  
\begin {equation}
\label {deg}
\left\{ \frac {\kappa}{2} { l}_{-1}^2 - 2 {l}_{-2}
+ \lambda U \right\} B = 0
. \end {equation}

Note that the definitions of $l_N$ and $U$ show easily that
for any $w= (w_0, w_1, \ldots)$ (not only in $V$),
$$ ([ l_N, U ]w)_n (x_1, \ldots, x_n) =
\sum_{J \subset \{1 , \ldots, n \}}
(l_N (T))_p ( x_J ) \times  w_{n-p} (x_{\{ 1, \ldots n\} \setminus J }) 
.$$
In order to compute $l_N (T)_p$, one
has to look at the Laurent expansion (when $x \to 0$) of
$T_{p+1} (x, x_1, \ldots, x_p)$.
Recall that $T_1 (x) = 1 / x^4$ and note that for $p \ge 1$,
\begin {equation}
\label {tp+1}
T_{p+1} (x, x_1, \ldots, x_p) = 2 x^{-2} T_p (x_1, \ldots, x_p) + o (x^{-2})
\end {equation}
(the only terms in the sum that contribute to the leading term
 are those corresponding to $x$ being visited 
first or last by the loop). 
It follows that $l_N T = 0$ if $N > 2$ and if $N=1$ (there are no $x^{-N-2}$ terms in the
expansion). Also, $l_2 T = (1,0, 0, \ldots )$ (the only case where there is an $x^{-4}$ 
term is $p+1=1$). Finally, 
$l_0 T = 2 T$ because of (\ref {tp+1}).
Hence,
$$
[l_N, U ] =  \left\{ 
 \begin {array}{l@{\hbox { if }}l}
 0 & N>2  \\
 \hbox {Id} & N=2  \\
 0 & N=1 \\
 2U & N= 0
 \end {array}
\right.$$
  
This enables as before to relate $\lambda$ to $\kappa$ and $h$: 
$$
l_2 ( \kappa l_{-1}^2 /2 - 2 l_{-2} ) B = l_2 ( - \lambda U B) =
- \lambda B - \lambda U l_2 B = - \lambda B
$$
and
$$
l_1 ( \kappa l_{-1}^2/2 - 2 l_{-2}) B =  l_1 ( - \lambda U  B) =
- \lambda U l_1 B =  0
.$$
This last relation implies that
$$
h = \frac {6 - \kappa}{2 \kappa}
$$
and the first one then shows that
$$
\lambda = (8- 3 \kappa) h = \frac {(8 - 3 \kappa)(6 - \kappa)}{2 \kappa},
$$
which are the formulae appearing in \cite {LSWr}.

This relation between  $h$ and $-\lambda$ is indeed that between
the highest-weight and the central charge for a representation of the
Virasoro algebra that is degenerate at level two.
Recall that if $\tilde l_n$'s are the generators of the Virasoro Algebra
and $C$ its central element, then 
$[ \tilde l_2, \tilde l_{-2} ] = 4 \tilde l_0 + C/2 $, so the little 
two by two linear system leading to the determination of $\kappa$ and $h$
for a degenerate highest-weight representation of the Virasoro algebra is
the same (and therefore leads to the same expression); roughly speaking, 
$l_{-2} - \lambda U /2$ plays the role of $\tilde l_{-2}$.
% i.e. one has to solve
%$$ \kappa \tilde l_{-1}^2   
%(recall that in the case of a representation of the 
%Virasoro algebra with central charge $c$, one has
%$[l_2, l_{-2}] = 4 l_0 + c/2 $).
% In other words, define
%for all $n \ge -2$,
%$$
%\tilde l_n = l_n -  \frac {\lambda U}{2} 1_{\{n=-2\}}.
%$$
%Then, for all $m,n \ge -2$,
%$$
%[ \tilde l_n , \tilde l_m ] = (n-m) 
%\tilde l_{n+m} + \frac {\lambda (n^3 -n)}{12} 1_{\{n=-m\}}
%$$
%when acting on $V$. Furthermore, 
%$\tilde l_n (B) = 0$ for $n \ge 1$, $\tilde l_0 B = h B$
%and $(\kappa \tilde l_{-1}^2 /2 - 2 \tilde l_{-2} ) B=0$.
%It follows that, just as for the degenerate representations of the 
%Virasoro algebra with central charge 
%$-\lambda$ that
%$
%-\lambda= 
%-{(8 - 3 \kappa)(6 - \kappa)}/ {2 \kappa}$ and
%that 
%the highest weight is $h= (6 - \kappa)/2 \kappa$.

Note that  the previous considerations involving the Brownian
bubbles is valid only in the range $\kappa \in (0, 
8/3]$ and therefore for $c \le 0$. This corresponds to the 
fact that two-sided 
restriction measures exist only for $h \ge 5/8$. In this 
case all functions $B^{(h)}_n$ are positive for all (real) 
values of $x_1, \ldots, x_n$.

\subsection {Analytic continuation}

In the representations that we 
have just been looking at, we considered simple 
operators acting on simple rational functions. All the results depend
analytically on $\kappa$ (or $h$). In other words, for all 
real $\kappa$ (even negative!), if one defines the functions 
$B_n^{(h)}$ recursively, the operators $l_n$, the vector $B^{(h)}$
 and the
vector space $V=V^{(h)}$  as before, then one obtains 
a highest-weight representation of ${\cal A}$ with 
highest weight $h$. 
The values of $\kappa$, $\lambda$ and $h$ are still related 
by the same formula, but do not correspond 
necessarily to a quantity that is directly  
relevant to the SLE curve or the restriction measures.

When $h \in (0, 5/8)$, the functions $B^{(h)}_n$ 
can still be interpreted as renormalized probabilities for
one-sided restriction measures.
They are therefore positive for all positive $x_1, \ldots, x_n$
but they can become negative for some negative values of the 
arguments. The ``SLE + bubbles'' 
interpretation of the 
degeneracy (i.e. of the relation (\ref{deg}))
is no longer valid since the ``density of bubbles''
becomes negative (i.e. the corresponding central charge is positive).
In this case, the local martingales measuring the effect 
of boundary perturbations are no longer bounded (and do not 
correspond to conditional probabilities anymore).

For negative $h$, the functions $B_n^{(h)}$ 
can still be defined. This time, the functions $B_n^{(h)}$ 
are not (all) positive, even when restricted on $(0, \infty)^n$
and they do not  correspond to any restriction measure. These facts correspond to 
``negative probabilities'' that are often implicit in the physics literature.

Note that $c$ (i.e. $- \lambda$) cannot take any value: For positive 
$\kappa$, $c$ varies in $(- \infty, 1)$ and for negative $\kappa$, it varies
in $[25, \infty)$.
The transformation $\kappa \leftrightarrow -\kappa$ corresponds to the
well-know $c \leftrightarrow 26-c$ duality (e.g. \cite {Ne}).

In other words, the $B_n^{(h)}$'s provide the
highest-weight representations
of ${\cal A}$ with highest weight $h$.
Each one is related to a highest-weight
representation of the Virasoro algebra that is degenerate at level 2.
Furthermore, all $B_n^{(h)}$'s are related by (\ref {sg}).

\section {Remarks}

In order to clarify the state of the art seen from a mathematical perspective,
let us now try to sum up things:

\begin {itemize}
\item
The interfaces of two-dimensional critical models 
(such as random cluster interfaces, that are very closely related 
to Potts models) are believed to be conformally invariant in the 
scaling limit. In some cases, this is proved (critical percolation,
uniform spanning trees). In some other cases (Ising, double-domino 
tiling), some partial results hold. Anyway, to derive conformal invariance,
it seems that one has to work on each specific model separately.

\item 
These interfaces can be constructed in a dynamic way i.e. they have a Markovian 
type property (at least the critical random cluster interfaces,
that have the same correlation functions as the Potts models).
Therefore, if conformal invariance holds, their scaling limit
{\em must} be one of the SLE curves. In general, these limits correspond to 
the SLE curves with $\kappa >4$ that are not simple curves. 
The correlation functions of the 2D statistical physics model are related to the 
fractal properties of the SLE curve, but the knowledge of the SLE curve 
is a much richer information than just the value of the exponents.

\item 
One can understand the dependence of the law of an SLE in a domain with 
respect to this domain via the restriction properties. This shows that some
specific ``finite-dimensional observables'' of the SLE curves satisfy some
relations. This can be reformulated in terms of highest-weight
representations  of the Lie 
algebra ${\cal A}$, and explains the 
relation between the physics models and these 
representations. Also, it makes it
possible to define conformal fields
via SLE.
However, and we think that this has to be again stressed, since the 
initial purpose was to understand the statistical physics models and their
behaviour, the SLE itself is a more natural way.
Also, one should also again emphasize that in the present paper, the ``correlation functions'' $B_n^{(h)}$ do correspond only 
indirectly with the curve $\gamma$ (via the cloud of 
Brownian bubbles) when the central charge does not vanish.

\end {itemize}

All functions described in the present paper deal with the boundary (or ``surface'') behaviour of the systems. One may want to develop a similar theory for points lying in the inside of the upper half-plane (``in the bulk''). Beffara's results \cite {Be}
 (for instance
in the case $\kappa=8/3$)  provide a first 
step in this direction, and show that the definition of these
correlation functions themselves is not 
an easy task.

\medbreak
\noindent
{\bf Acknowledgements.}
Thanks are of course due to Greg Lawler and Oded Schramm,
in particular because of the instrumental role played by the ideas
developed in the paper \cite {LSWr}.
We have also benefited from 
very useful discussions with Vincent Beffara and Yves Le Jan.
R.F. acknowledges support and hospitality of IHES.

------------------

Laboratoire de Math\'ematiques

Universit\'e Paris-Sud

91405 Orsay cedex, France

emails: rolandf@ihes.fr, wendelin.werner@math.u-psud.fr
\end {document}